\documentclass[aps,nofootinbib,showpacs,preprintnumbers,letterpaper,twocolumn,superscriptaddress,prl]{revtex4-1}

\usepackage{amsmath,amssymb}
\usepackage{graphicx}
\usepackage{bm}
\usepackage{color}
\usepackage{ulem}
\usepackage{epstopdf}

\newcommand{\E}{\ensuremath{\mathrm{e}}}
\newcommand{\I}{\ensuremath{\mathrm{i}}}

\newcommand{\ket}[1]{| #1\rangle}

\newcommand{\ev}[1]{\langle #1 \rangle}

\newcommand{\gbf}[1]{\boldsymbol #1}

\begin{document}
 
\title{Dimerized Solids and Resonating Plaquette Order in SU($N$)-Dirac Fermions}

\author{Thomas C. Lang}
\email{tcl@bu.edu}
\affiliation{Department of Physics, Boston University, Boston, MA 02215, USA}
\author{Zi Yang Meng}
\affiliation{Center for Computation and Technology, Department of Physics and Astronomy, Louisiana State University, Baton Rouge, Louisiana 70803, USA}
\author{Alejandro Muramatsu}
\affiliation{Institut f\"{u}r Theoretische Physik III, University of Stuttgart, 70550 Stuttgart, Germany}
\author{Stefan Wessel}
\affiliation{Institut f\"ur Theoretische Festk\"orperphysik, JARA-FIT and JARA-HPC, RWTH Aachen University, 52056 Aachen, Germany}
\author{Fakher F. Assaad}
\affiliation{Institute for Theoretical Physics and Astrophysics, University of W\"urzburg, 97074 W\"urzburg, Germany}

\begin{abstract}
We study the quantum phases of fermions with an explicit SU($N$)-symmetric, Heisenberg-like nearest-neighbor flavor exchange interaction on the honeycomb lattice at half filling. Employing projective (zero temperature) quantum Monte Carlo simulations for even values of $N$, we explore the evolution from a weak-coupling semimetal into the strong-coupling, insulating regime. Furthermore, we compare our numerical results to a saddle-point approximation in the large-$N$ limit. From the large-$N$ regime down to the SU(6) case, the insulating state is found to be a columnar valence bond crystal, with a direct transition to the semimetal at weak, finite coupling, in agreement with the mean-field result in the large-$N$ limit. At SU(4) however, the insulator exhibits a subtly different valence bond crystal structure, stabilized by resonating valence bond plaquettes. In the SU(2) limit, our results support a direct transition between the semimetal and an antiferromagnetic insulator.
\end{abstract}

\pacs{02.70.Ss,71.10.Fd,71.10.Hf,71.27.+a,73.43.Nq}
%see http://www.aip.org/pacs/pacs2010/individuals/pacs2010_regular_edition/alpha_index.html
% 02.70.Ss Monte Carlo methods, quantum Monte Carlo
% 71.10.Fd Lattice fermion models (Hubbard model, etc.)
% 71.10.Hf Non-Fermi-liquid ground states, electron phase diagrams and phase transitions in model systems
% 71.27.+a Strongly correlated electron systems
% 73.43.Nq Quantum phase transitions (see also 64.70.Tg Quantum phase transitions in equations of state, phase equilibria and phase transitions)

\maketitle
In dealing with quantum field theories or quantum statistical systems, perturbative expansions in the couplings may become unreliable due to ultraviolet or infrared singularities. A different route to explore the region of strong interactions was proposed by 't Hooft long ago, based on enlarging the number of internal degrees of freedom of a theory, eventually leading to a large-$N$ expansion~\cite{Hooft74}. In particular, in the context of strongly correlated electronic systems, large-$N$ theories paved the way towards a controlled theoretical understanding of new states of matter, such as the flux phase \cite{Affleck88,Marston89} or spontaneously dimerized two-dimensional spin systems \cite{Read89,Read90}. Although large-$N$ theories may have appeared as pertaining to a purely theoretical domain, in recent years, novel interest arose in the physics of correlated fermions with an exact SU($N$) flavor exchange symmetry. In fact, such systems can be realized in ultracold fermionic alkali- and alkaline-earth atoms in optical lattices~\cite{Wu03,Wu03,Honerkamp04,Jordens08,Schneider08,Cazalilla09,Gorshkov10,Wu10}, where significant experimental progress has been reported~\cite{experimentalprogress}, as well as in quantum dot arrays~\cite{Onufriev99} and at special points of coupled spin-orbital systems~\cite{Kugel73}. SU($N$) systems represent hot candidates to realize Mott insulators with fermionic cold atoms in optical lattices~\cite{Hazzard12,Bonnes12,Messio12}. Various theoretical approaches have been employed to study phases which emerge from an enhanced SU($N$) symmetry such as unconventional antiferromagnets~\cite{Toth10,Corboz12,Corboz11}, generalized valence bond solids \cite{Corboz07,Arovas08,Hermele11,Corboz12,Song13}, algebraic and chiral flavor liquids~\cite{Affleck88,Hermele05,Assaad05,Xu10,Corboz12,Cai12,Hermele11,Song13,Hermele09,Szirmai11} and others~\cite{RappManmanaCapponi}, as well as the quantum phase transitions between them~\cite{Honerkamp04,Assaad05,Hermele05,Wu06,Beach09,Daley08,Hermele09,Cazalilla09,Klingschat10,Toth10,Xu10,Gorshkov10,Hazzard12,Cai13,Cai12,Corboz12,Corboz13,Corboz11,Wang13}. 

Of particular interest in the search for exotic quantum phases is the physics that arises from SU($N$)-symmetric flavor exchange mechanisms in the strong correlation regime. While the large-$N$ domain can be accessed in terms of a $1/N$ expansion, accounting for Gaussian fluctuations around the mean-field saddle point, the physical realizations still reside within the regime of low to intermediate values of $N$, where some of the possible exotic states have indeed been identified~\cite{Assaad05,Hermele05,Wu06,Beach09,Klingschat10,Toth10,Corboz11,Corboz12,Corboz13}. Hence, an unbiased assessment of the validity of the large-$N$ results down to the lowest possible values becomes very important in light of these recent developments.

Here, we employ numerically exact quantum Monte Carlo (QMC) simulations to explore the physics of such fermionic systems. In particular, we consider a model of fermions with $N$ flavors, coupled via a nearest-neighbor SU($N$)-symmetric flavor exchange. The Hamiltonian that describes this system is given by
\begin{eqnarray}
 H&=&- t \sum_{ \langle i,j\rangle,\alpha } (c^\dagger_{i\alpha} c^{}_{j\alpha} + \mathrm{H.c.}) \nonumber\\
  & &-\frac{J}{2N} \sum_{ \langle i,j\rangle,\alpha,\beta} (c^\dagger_{i\alpha} c^{}_{j\alpha}c^\dagger_{j\beta} c^{}_{i\beta} + c^\dagger_{j\alpha} c^{}_{i\alpha}c^\dagger_{i\beta} c^{}_{j\beta}).
\end{eqnarray}
Here, $t$ denotes the tunneling amplitude of the $N$-flavor fermions with creation (annihilation) operators $c^\dagger_{i\alpha}$ ($c_{i\alpha}$) of flavor $\alpha=1,...,N$ on lattice site $i$. Furthermore, $J$ sets the strength of the nearest-neighbor flavor exchange interaction. We consider the system at half filling, $N_\text{f}/N_\text{s}=N/2$, where $N_\text{f}$ denotes the total number of fermions on a lattice with $N_\text{s}$ sites, for which unbiased quantum Monte Carlo simulations can be performed without a sign problem. If the Hilbert space were further constrained to exactly $N/2$ particles on each lattice site, $H$ would reduce to an exact SU($N$) symmetric Heisenberg model $H_\text{H}=J/2N \sum_{ \langle i,j\rangle} \mathbf{S}_i \cdot \mathbf{S}_j$, with $\mathbf{S}_i$ being the vector of the SU($N$) spin operators $S^a_i=\sum_{\alpha,\beta} c^\dagger_{i\alpha} T^a_{\alpha\beta} c^{}_{i\beta}$, expressed in terms of the generators $T^a$, ${a=1,\ldots,N^2-1}$ of SU($N$) in the fundamental representation, with $\mathrm{Tr}(T^a T^b)=\delta_{ab}/2$ [e.g., $T^a=\sigma^a/2$ in terms of the Pauli matrices $\sigma^a$ for SU(2)]. This would result, e.g., in the large $U$-limit of a model that in addition to $H$ also includes a local Hubbard-$U$ interaction, which reduces the local particle fluctuations around the mean value of $N/2$. In the large-$N$ limit, and considering only the paramagnetic saddle point, these fluctuations become irrelevant and the Hubbard-$U$ merely fixes the average particle number to $N/2$. 
The Hamiltonian in Eq.~(1) indeed equals the $U=0$ limit of the Hubbard-Heisenberg Hamiltonian of the seminal works in Refs.~\onlinecite{Read89,Affleck88,Marston89}, and represents an unrestricted SU($N$)-symmetric ${t\mbox{-}J}$ model. It has been considered previously using QMC on the square lattice, where an exotic gapless spin liquid was obtained for $N=4$ flavors~\cite{Assaad05}. In the following, we consider the case of the honeycomb lattice, motivated also by recent studies for the SU($2$) Hubbard model~\cite{HubbardHoneycomb}, and are in particular interested in the response of the weak-coupling SU($N$) semimetal (SM) to an explicit SU($N$)-symmetric flavor exchange interaction $J$.

\begin{figure}[t]
   \centering
   \includegraphics[width=0.95\columnwidth,clip=true]{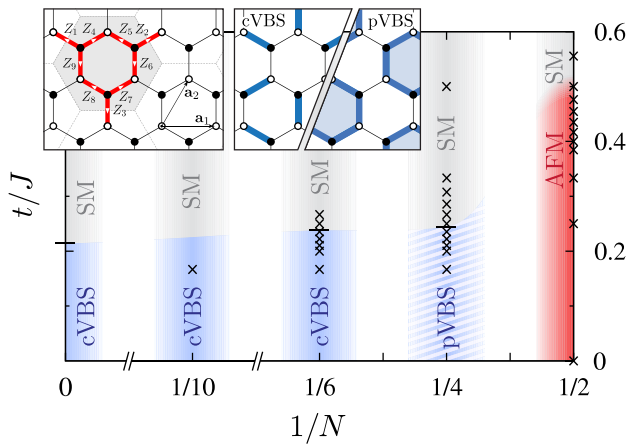}
\caption{Ground state phase diagram of fermions with SU($N$)-symmetric flavor exchange on the honeycomb lattice. Crosses denote parameters at which QMC simulations have been carried out. For all considered (even) $N$ the system undergoes a quantum phase transition from a semimetal (SM) to an insulator. For ${N\ge 6}$ the insulating state is a columnar valence bond solid (cVBS), while at ${N = 4}$ it is a VBS with resonating valence bond plaquettes (pVBS); both are depicted in the right inset. At ${N=2}$ an antiferromagnetic insulator (AFM) appears. The left inset shows the lattice structure with the six-sites unit cell employed in the large-$N$ calculations.
   \label{fig:pd}}
\end{figure}

\textit{Large-$N$.} Before presenting QMC results, we consider the mean-field decoupling of $H$ in terms of the bond mean-fields $\chi_{ij}=|\chi_{ij}|\E^{\I\phi_{ij}}=\langle\sum_\alpha c^\dagger_{i\alpha} c^{}_{j\alpha}\rangle/N$, which becomes exact in the large-$N$ limit. The $\chi_{ij}$ carry a phase $\phi_{ij}$, and $|\chi_{ij}|^2$ relates the bond strength $\langle \mathbf{S}_i \cdot \mathbf{S}_j\rangle$. We numerically solve the mean-field equations self-consistently for a six-site unit cell (with nine bonds), which retains the full lattice symmetry (cf. the left inset of Fig.~\ref{fig:pd}). This leads to the following phase diagram: At large $t/J$, the kinetic energy dominates and all the $\chi_{ij}$ are equal and real, thus the system in this region is a flux-less SM. Below a critical value near ${t/J = 0.21}$, the system undergoes a continuous quantum phase transition into a columnar valence bond solid (cVBS) phase with a Kekule pattern~\cite{Read90,Hou07}, illustrated in the right inset of Fig.~\ref{fig:pd}. For comparison, we note that on the square lattice, the noninteracting Fermi sea is unstable, and in the large-$N$ limit d-density wave states occur immediately at weak coupling, while a VBS with box dimerization emerges at large exchange coupling~\cite{Assaad05}. 

\begin{figure}[t]
   \centering
   \includegraphics[width=0.90\columnwidth,clip=true]{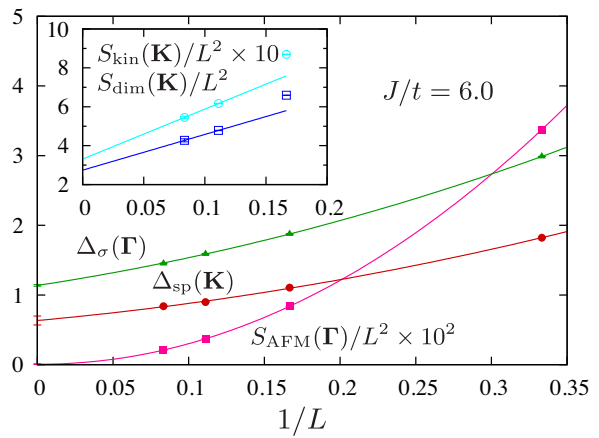}
   \caption{${N = 10}$ --- Finite size extrapolation of the AFM structure factor ${S_{\text{AFM}}(\gbf{\Gamma})}$, the spin gap ${\Delta_{\sigma}(\gbf{\Gamma})}$ and the single-particle gap ${\Delta_{\text{sp}}(\mathbf{K})}$ in the VBS for SU(10) at ${J/t=6.0}$. The inset shows the finite size extrapolation of the structure factors $S_\text{dim}(\mathbf{K})$ and $S_\text{kin}(\mathbf{K})$.
   \label{fig:SU10}}
\end{figure}

\textit{QMC method}. To explore the phase diagram beyond the large-$N$ limit, and to assess the stability range of the large-$N$ results, we employ a SU($N$)-generalized formulation of the projector QMC~\cite{Sugiyama86,AsEv08,Assaad05}, which allows for the numerically exact evaluation of ground state properties for all even values of $N$. Observables are obtained as $\langle \Psi_0 | O |\Psi_0 \rangle = \lim_{\Theta \rightarrow \infty} \langle \Psi_\text{T} |\mbox{e}^{-\Theta H/2} O \mbox{e}^{-\Theta H/2} | \Psi_\text{T} \rangle / \langle \Psi_\text{T} |\mbox{e}^{-\Theta H} | \Psi_\text{T} \rangle$. We use a trial wave function ${\ket{\Psi_\text{T}} = \prod_{\alpha} \ket{\Psi_\text{T}}_{\alpha}}$, where $\ket{\Psi_\text{T}}_{\alpha}$ is the ground state of the single particle Hamiltonian ${H_{\alpha}^0 = -t \sum_{\langle i,j\rangle} c^{\dagger}_{i\alpha} c_{j\alpha} \exp\big(\frac{2\pi\I}{\Phi_{0}}\int_{\mathbf{r}_i}^{\mathbf{r}_j}\mathrm{d}{\boldsymbol\ell}\cdot\mathbf{A}\big) + \mathrm{H.c.} }$ in the flavor $\alpha$ Hilbert subspace, where ${\Phi_{0} = h e/c}$ denotes the flux quantum, and $\mathbf{r}_i$ the position of lattice site $i$. The flux ${\Phi/\Phi_0}=10^{-4} $ is chosen sufficiently small to lift the ground state degeneracy in ${|\Psi_\text{T}\rangle}$. We performed QMC simulations on finite systems of linear extent $L$ and $N_\text{s}=2L^2$ sites, with periodic boundary conditions. Projection parameters ${\Theta t = 30}$ and an imaginary time discretization of ${\Delta\tau t = 0.05}$ were found sufficient to obtain converged ground-state quantities within statistical uncertainty. From a fit of the imaginary-time displaced Green's function~\cite{Feldbacher01} $G(\mathbf{q},\tau) = \ev{\frac{1}{2N}\sum_{s,\alpha} c_{\mathbf{q}s\alpha}^{\dagger}(\tau) c_{\mathbf{q}s\alpha}(0)}$ to its long-time behavior, $\lim_{\tau\to\infty} G(\mathbf{q},\tau)\propto\E^{-\tau\Delta_{\text{sp}}(\mathbf{q})}$, the single-particle gap $\Delta_{\text{sp}}=\Delta_{\text{sp}}(\mathbf{K})$ can be extracted without an analytical continuation. Here, the momentum $\mathbf{q}$ is defined with respect to the coordinates of the two-site unit cells of the honeycomb lattice that form a triangular lattice, $\mathbf{K}$ denotes a corner of the hexagonal Brillouin zone (where the Dirac points of the SM reside), and $s=A,B$ corresponds to the site of the unit cell that belongs to sublattice $A$ and $B$, respectively. Similarly, we obtain the spin gap ${\Delta_{\sigma}(\gbf{\Gamma})}$ from the time-displaced spin-spin correlation function in the antiferromagnetic (AFM) sector, $S_{\text{AFM}}(\gbf{\Gamma},\tau) = \frac{1}{N_\text{s}}\sum_{i,j} \epsilon_i\epsilon_j \ev{\mathbf{S}_i(\tau) \cdot \mathbf{S}_j(0)}$, where $\epsilon_i=\pm 1$ if site $i$ belongs to sublattice $A$ ($B$).
\begin{figure}[t]
   \centering
   \includegraphics[width=0.85\columnwidth,clip=true]{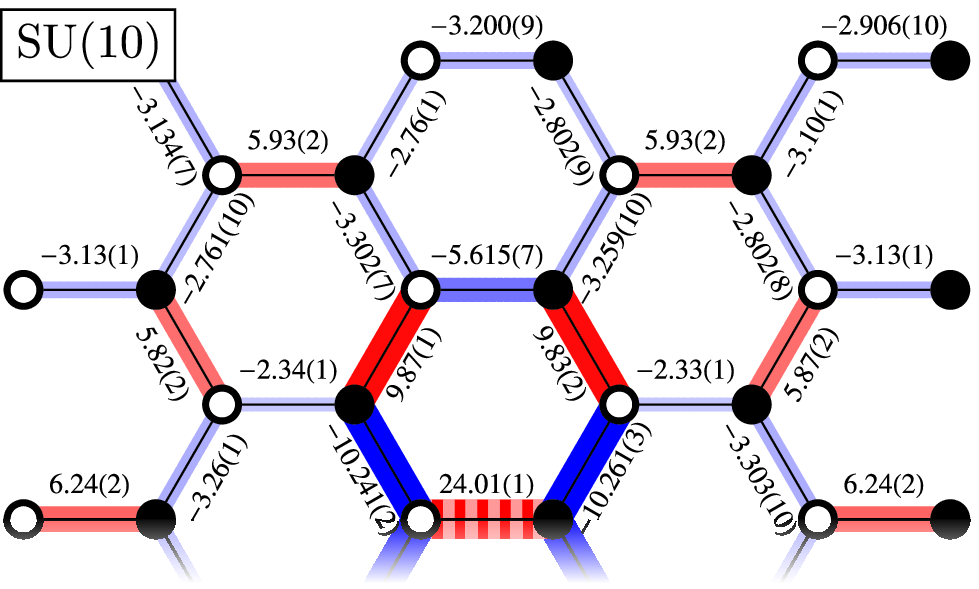}
   \caption{${N = 10}$ --- Real-space dimer correlations $D_{ij,kl}$ on a ${L=6}$ lattice for SU(10) at ${J/t=6.0}$, with the striped reference bond. $D_{ij,kl}$ is indicated by colors and the line thickness in addition to the explicit values given for each bond.
   \label{fig:dimdim}}
 \vspace{1em}
   \includegraphics[width=0.9\columnwidth,clip=true]{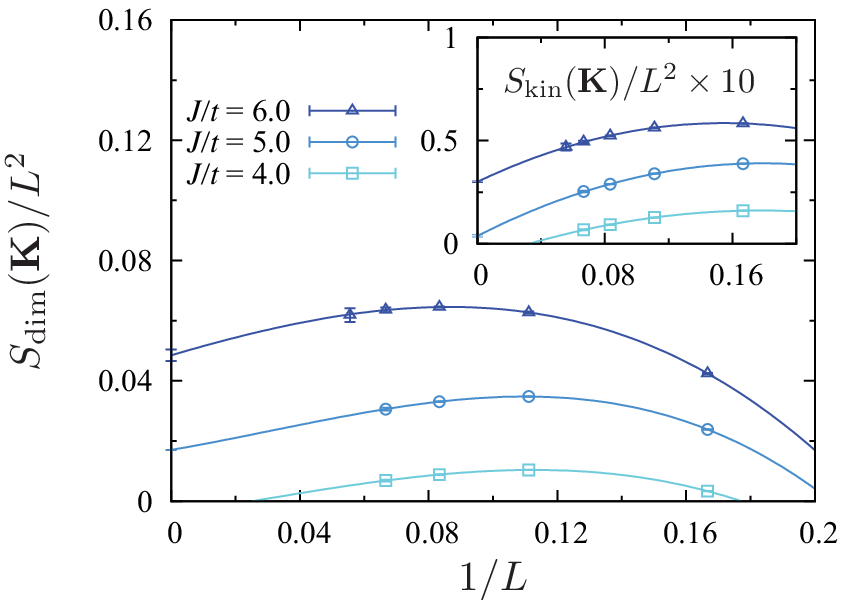}
   \caption{${N = 4}$ --- Finite size scaling of the dimer and kinetic structure factors $S_\text{dim}(\mathbf{K})$ and $S_\text{kin}(\mathbf{K})$ at SU(4).
   \label{fig:dimCorrSU4}}
\end{figure}
The equal time value $S_{\text{AFM}}(\gbf{\Gamma})=S_{\text{AFM}}(\gbf{\Gamma},\tau=0)$ provides the structure factor for long-range AFM order on this bipartite lattice. In order to determine the dimerization pattern of the VBS phase stabilized in the large-$N$ analysis, we measure the SU($N$) dimer correlation function $D_{ij,kl}=\langle\!\langle (\mathbf{S}_i \cdot \mathbf{S}_j) (\mathbf{S}_k \cdot \mathbf{S}_l)\rangle\!\rangle$, where $\langle\!\langle O_1, O_2 \rangle\!\rangle= \langle O_1 O_2 \rangle - \langle O_1 \rangle \langle O_2 \rangle$ denotes a cumulant. From the Fourier transformation of $D_{ij,kl}$ for a set of parallel bonds $\langle ij\rangle$, $\langle kl\rangle$, we furthermore obtain a corresponding structure factor $S_\text{dim}(\mathbf{K})$. Similarly, we define a structure factor $S_\text{kin}(\mathbf{K})$ from the kinetic energy correlators $\langle\!\langle c^\dagger_i c^{}_j+ c^\dagger_j c^{}_i, c^\dagger_k c^{}_l+ c^\dagger_l c^{}_k\rangle\!\rangle$ among a set of parallel bonds of the honeycomb lattice.

\begin{figure}[t]
   \centering
   \includegraphics[width=0.9\columnwidth]{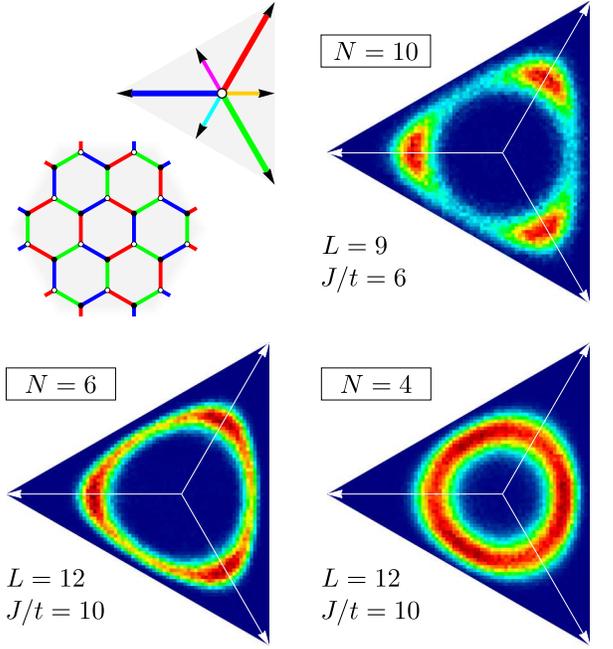}
   \caption{Histograms of the dimer density along three bond types (illustrated in the top left panel), which correspond to three realizations of the same columnar VBS. Predominant weight at the corners in the phase space represent the columnar states as seen at $N=10$ and 6. A resonating plaquette is formed on two different bond types and is signaled by weight at the midpoint between two bond directions as realized at SU(4). All histograms have been rescaled in order to ease the identification of their characteristic distributions.
   \label{fig:vbs}}
\end{figure}

\textit{QMC results}. As shown in the phase diagram in Fig.~1, our findings for $N\geq4$ essentially agree with the scenario obtained from the large-$N$ analysis, with the critical ratio $t/J$ of the SM-VBS transition slightly increasing with $1/N$. For example, from the data at ${J/t=6}$ for $N=10$, shown in Fig.~\ref{fig:SU10}, a robust finite single-particle $\Delta_{{\text{sp}}}(\mathbf{K})$ and spin-gap $\Delta_{\sigma}(\gbf{\Gamma})$ is extracted. The order parameters derived from the characteristic VBS structure factor $S_\text{dim}(\mathbf{K})$, as well as from the kinetic energy-based structure factor $S_\text{kin}(\mathbf{K})$ extrapolate to a finite value in the thermodynamic limit (TDL), as shown in the inset of Fig.~\ref{fig:SU10}, while the AFM order parameter $S_{\text{AFM}}/N_\text{s}$ vanishes in the TDL, as expected for the cVBS phase. Figure~\ref{fig:dimdim} shows the corresponding dimer correlations in real space on a $L=6$ lattice. The robust characteristic Kekule pattern (cf. the right inset in Fig.~\ref{fig:pd}) generates a mass term without breaking the sublattice symmetry. Furthermore translation symmetry of the original lattice is broken: the unit cell contains six orbitals, and forms a triangular lattice.

Decreasing $N$ down to $6$, the cVBS order is increasingly weakened but remains persistent; we observed similar finite size scaling behavior of observables at SU(6) as those at SU(10). The situation changes at SU(4), as shown in Fig.~\ref{fig:dimCorrSU4}, where we observe an unusual finite size scaling behavior of $S_\text{dim}(\mathbf{K})$ (main panel) and $S_\text{kin}(\mathbf{K})$ (inset). While the extrapolation still yields long-range dimer correlations at strong coupling ($J/t\gtrsim4.3$) in the TDL, it hints at a possible different VBS pattern than the cVBS found at higher $N$. 

While the dimer correlation pattern (e.g., Fig.~\ref{fig:dimdim}) allows to identify VBS order of a certain momentum, it cannot distinguish between VBS states of different structures, such as cVBS and plaquette VBS (pVBS) states, as they relate to the same wave vector $\mathbf{K}$. At SU(4), fluctuations could possibly be sufficiently strong to soften the cVBS order by forming resonating valence bond plaquettes -- the antisymmetric combination of the two singlet coverings around a hexagonal plaquette -- which order into a $\sqrt{3}\times\sqrt{3}$ pVBS, illustrated in the right inset of Fig.~\ref{fig:pd}.

In order to unveil a possible change in the nature of the VBS, we employ a histogram technique~\cite{Albuquerque11}: we track the contributions to the dimer correlations originating from the three different bond types defined by red, green and blues bonds in Fig.~\ref{fig:vbs}. Histograms with a dominant weight along the three main bond directions indicate cVBS order. Indeed, the data in Fig.~\ref{fig:vbs} for both SU(10) and SU(6) exhibit peaks at the axial corners, representing the Kekule pattern. In the TDL one of the three peaks for the equivalent realizations will prevail, as the translational symmetry of the lattice is spontaneously broken. In case two bond types contribute equally well, and resonating valence bond plaquettes are formed, the histogram shows predominant weight at the midpoint between two of the major bond directions. Indeed, this can be readily identified in the histogram for SU(4) in Fig.~\ref{fig:vbs}, such that the transition from the cVBS at SU(6) to the pVBS at SU(4) can be unambiguously identified. Interestingly, while both VBS orders break the same lattice symmetry they can be distinguished by the dimer histogram \cite{VBSorder,Corboz13}.

\begin{figure}[t]
   \centering
   \includegraphics[width=\columnwidth]{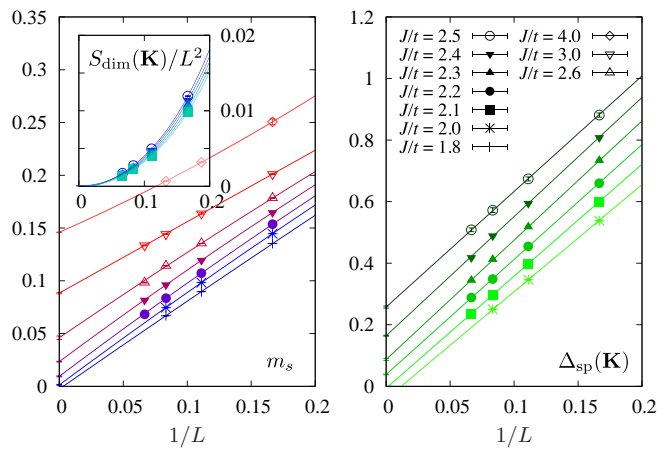}
   \caption{${N = 2}$ --- Staggered magnetization ${m_s = \sqrt{S_\text{AFM}(\gbf{\Gamma})}/L}$ and single-particle gap $\Delta_{\text{sp}}(\mathbf{K})$ as functions of the system size for SU(2) and different ratios $J/t$. The inset shows the finite size scaling of the structure factor $S_\text{dim}(\mathbf{K})$ inside the range ${2.1 \lesssim J/t \lesssim 2.5}$.
   \label{fig:SU2spgapmag}}
\end{figure}

While we thus find the large-$N$ scenario to describe well the region $N\geq6$, we observe its partial breakdown at $N=4$, where the structure of the VBC changes. Furthermore, in the SU(2) case, the ground state becomes a N\'{e}el antiferromagnet (AFM) for ratios ${J/t \approx2.1}$. This can be seen from the finite size data of the staggered magnetization $m_s(L) \equiv \sqrt{S_\text{AFM}(\gbf{\Gamma})}/L$ in the left panel of Fig.~\ref{fig:SU2spgapmag}. With the available system sizes, the data are consistent with a direct SM-AFM transition, as the opening of the single-particle gap and the onset of AFM coincide. In fact, as seen from the right panel of Fig.~\ref{fig:SU2spgapmag}, a finite single-particle gap in the TDL opens beyond ${J/t\approx 2.1}$. In contrast to the cases of larger $N$, we do not observe VBS formation at SU(2) as seen from the rapid finite-size downscaling of the corresponding structure factor $S_\text{dim}(\mathbf{K})$ shown in the inset of Figs.~\ref{fig:SU2spgapmag} in comparison with Figs. \ref{fig:SU10} and \ref{fig:dimCorrSU4}. At ${N=2}$, one would expect the transition between the SM and AFM to be similar to that for the $\pi$-flux Hubbard model and honeycomb lattice \cite{HubbardHoneycomb,Chang12}. Larger system sizes and more precise simulations will be required to clarify this point. 

We explored the phase diagram of fermions with an SU($N$)-symmetric flavor exchange interaction on the honeycomb lattice at half filling. Our quantum Monte Carlo simulations confirm that the scenario from the large-$N$ approach holds down to $N=6$. This result is encouraging for the general large-$N$ approach, in light of its application to cold gases of alkaline earth metals, where $N$ can indeed take on large values (e.g., $N=6$ for ${}^{173}$Y and $N=10$ for ${}^{87}$Sr). In the context of the present flavor exchange model, where fluctuations are enhanced by the charge dynamics, the strong coupling region at $N=2$ and 4 however deviates from the large-$N$ limit. Based on quantum Monte Carlo simulations, we exhibited the presence of resonating valence bond plaquettes at SU(4), separating the antiferromagnetic strong coupling phase at $N=2$ from the Kekule ordered region for $N \geq 6$. In would be interesting for future research, to address the nature of the quantum phase transitions that separate these three different large-$J$ regions when varying the flavor exchange symmetry $N$, e.g., within a feasible continuous-$N$ generalization of the model. This would offer the prospects of a potential candidate for a deconfined quantum phase transition \cite{Senthil04} between the AFM and pVBS as well as the possibility to study the proliferation of vortices in the Kekule structure which trigger a phase transition from cVBS to the pVBS phase.

We thank L.~Balents, M.~Hermele, A.~M.~L\"{a}uchli and S.~Sachdev for helpful comments and discussions. This research was supported in part by EPSCoR Cooperative Agreement EPS-1003897 (ZYM), DFG AS102/4-3 (FFA), DFG WE 3649/3-1 (SW) and DFG FOR1807 (FFA, SW). Furthermore, we acknowledge the JSC J\"ulich, JARA-HPC, the HLRS Stuttgart and the LRZ-Munich for the allocation of CPU time.

\end{document}